\documentclass[%
reprint, 
longbibliography,
superscriptaddress,
nofootinbib,
amsmath,amssymb,
aps,
prx,
]{revtex4-1}

\usepackage{graphicx,amsmath,amssymb, color}

\usepackage{bm}
\usepackage{hyperref}
\usepackage{physics}
\usepackage{quantikz}

\def\equationautorefname~#1\null{Eq. (#1)\null}

\newcommand{\appref}[1]{\hyperref[#1]{App.~\ref*{#1}}}

\renewcommand\vec{\bm}

\usepackage{bbold}

\begin{abstract}
    Using a 36-qubit quantum processor, we demonstrate that, by operating in conjunction with a classical machine learning agent, quantum computers can discover instances of interesting quantum many-body dynamics. The central object in this new mode of use of a quantum device is an ``interest function'' defined for a given circuit (family) instance that can be evaluated on a quantum computer. The circuit is adapted by the learning agent to maximize interest. We illustrate this approach using two examples and show that, within a sufficiently general circuit family, two simple interest functions based on (i) binary classifiability of evolved states and (ii) spectral properties of the unitary circuit, are maximized by discrete time crystals (DTCs) and dual-unitary circuits, respectively. For the classifiability-based interest function, we implement the protocol on a superconducting quantum processor and find that it indeed discovers DTCs with high probability. For the dual-unitaries, our simulations of the dynamics suggest that an interest-function optimization would have set us close to a discovery of such unitaries. Our results using quantum devices and accompanying simulations suggest that learning agents with access to quantum-computing resources can almost autonomously discover new phenomena in many-body quantum dynamics, and establish the design of good interest functions optimizable in hybrid devices as a paradigm for quantum many-body physics.
\end{abstract}
\begin{document}

\title{Running Quantum Computers in Discovery Mode}
\author{Aydin Deger}
\thanks{These authors contributed equally and are listed in alphabetical order.}
\affiliation{Department of Physics, Clarendon Laboratory, University of Oxford, Oxford OX1 3PU, United Kingdom}
\author{Benedikt Placke}
\thanks{These authors contributed equally and are listed in alphabetical order.}
\affiliation{Rudolf Peierls Centre for Theoretical Physics, University of Oxford, Oxford OX1 3PU, United Kingdom}
\author{G. J. Sreejith}
\thanks{These authors contributed equally and are listed in alphabetical order.}
\affiliation{Indian Institute of Science Education and Research, Pune MH 411008, India }
\author{Alessio Lerose}
\affiliation{Rudolf Peierls Centre for Theoretical Physics, University of Oxford, Oxford OX1 3PU, United Kingdom}
\affiliation{Institute for Theoretical Physics, KU Leuven, Celestijnenlaan 200D, 3001 Leuven, Belgium}
\author{S.~L. Sondhi}
\affiliation{Rudolf Peierls Centre for Theoretical Physics, University of Oxford, Oxford OX1 3PU, United Kingdom}

\maketitle

\begin{figure*}
    \centering
    \includegraphics{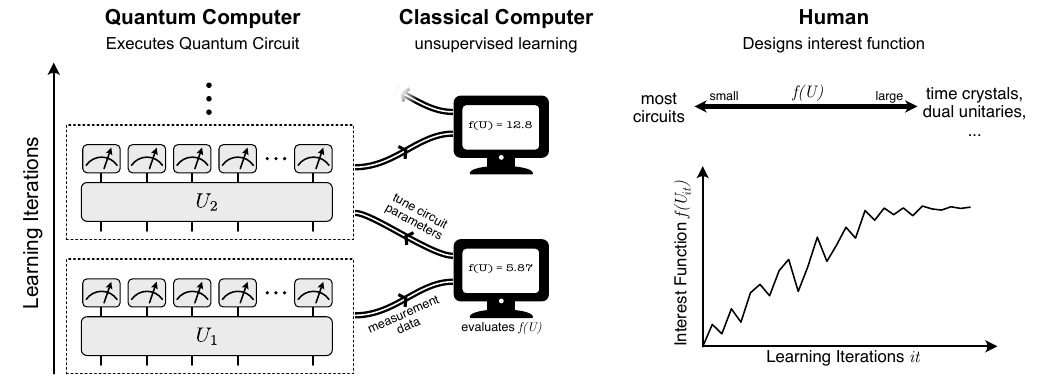}
    \caption{{\bf Running Quantum Computers in discovery mode.} The central idea is to use a quantum computer to run a parameterized circuit and then use unsupervised machine learning to steer the circuit towards realizing interesting physics. The central human input is to provide the definition of what is deemed ``interesting''. More precisely, one has to define a so-called ``interest function'' which is then maximized using classical learning algorithms, and which ideally discriminates most circuits from those that we find interesting. In this work, we provide two specific examples of such functions, and show numerically that these would have naturally led to the discovery of discrete time crystals and dual-unitary circuits, respectively. In a scalable implementation, learned parameters and representations may be transferred as the system size and circuit family are progressively enlarged, rather than restarting the search from random circuits at each scale.}
    \label{fig:toc}
\end{figure*}

\section{Introduction}

The project of creating a quantum computer involves, at a minimum, building quantum devices with increasing numbers of qubits acted upon by quantum circuits of increasing depth. As of today devices executing circuits of depth 1000 acting on 1000 qubits seem within reach. Such ``physical megaquops'' \cite{preskill2025megaquop} devices would be able to realize a mind-boggling number of distinct quantum circuits, roughly $10^{10^{5.8}}$ by a naïve count.\footnote{While we hasten to note that they will not all lead to physically distinct unitary evolutions, they still leave room for an extraordinary number of such evolutions.} In the face of such numbers, an increasingly important question is how to utilize such devices. Most proposals to date fall roughly into two categories: one can either make use of the device as a universal digital quantum computer, ideally in a setting where quantum error correction ensures flawless hardware operation, or use it as a platform for analogue quantum simulation to investigate the dynamics of quantum many body systems of interest. Here, we propose a third usage which we term ``discovery mode''. Running a device in discovery mode means searching over the space of possible quantum circuits to find interesting instances. This approach focuses our attention on defining what circuits might be considered interesting, as well as on establishing procedures to identify them. 

To make this more concrete, the first part of our proposal involves identifying what we call ``interest functions'', which will depend jointly on a certain number of initial states and the results of a certain number of final measurements, both bracketing the operation of a given unitary circuit. It is these functions that really incorporate the idea of what is ``interesting'', and we give two very different examples below.
The second part of the proposal would involve using a classical learning agent to explore the space of circuits with the goal of finding at least a local maximum of the interest function. 

In bringing these components together, our primary aim is to leverage the computational power of quantum platforms to handle the quantum dynamics efficiently. At the same time, we can utilize significantly larger quantum platforms to generate substantial amounts of data, which can be analyzed in tandem with classical computers. This integrated approach — using classical learning algorithms alongside quantum platforms as summarized in \autoref{fig:toc} — aligns with the growing evidence that such algorithms will succeed in the long run, as demonstrated most notably by recent advancements in artificial intelligence \cite{sutton2019bitter, yousefi2024learning}.

The broader program raises two distinct questions. First, what might useful interest functions look like? Second, can they be optimized over sufficiently rich families of circuits? The last question is sharpened by barren-plateau results for variational quantum algorithms, which show that generic random initialization in sufficiently expressive circuit families can yield an exponentially weak optimization signal, as well as by the glassy landscapes encountered in quantum annealing \cite{cerezo2021variation_review,peruzzo2014variational,mcclean2018barren,ortiz2021barren,Kadowaki1998Quantum,Albash2018Adiabatic,Santoro2002Theory,Altshuler2010Anderson,Bapst2013The}. The general interest functions are not operator expectation values unlike the cost functions encountered in VQAs and may offer a large trainable region. These results nevertheless caution against beginning with an unrestricted full-scale search. The lesson we draw from classical machine learning is instead to organize the search as a curriculum: begin with smaller systems and restricted, shallow, or parameter-shared circuit families; transfer the learned parameters to progressively larger systems; and only gradually enlarge the circuit class. In this division of labour, the quantum processor performs the irreducibly quantum task of generating data from dynamics beyond classical simulation, while the classical agent tackles a structured optimization problem and can benefit from continuing advances in representations, initialization, and search algorithms. Scalable discovery mode therefore requires interest functions for which both the measurement signal and the optimization signal remain accessible.

In this paper, we report a first study using a superconducting quantum processor interacting with a classical learning scheme in which we address these questions. In a first example, we utilize unsupervised learning algorithms to quantify the classifiability of states in a time series of states obtained from repeated applications of a finite-depth unitary circuit. We show that maximization of this interest function steers a suitable circuit family towards the realization of a discrete time crystal. We demonstrate the success of this approach using 12-36 qubit quantum devices that implement the parametrized circuit class to be optimized. To demonstrate the robustness of the strategy, we use classical simulations in more complex families of circuits.
In a second example, we perform classical simulations of a quantum device searching for unitary circuits with extremal spectral properties by using Landau functionals of unitaries as interest functions, and provide strong evidence that in this case as well the interest landscape is (at least locally) concave and that variational optimization would steer the circuit towards dual-unitarity. 

Together, our two examples address the questions raised before, and show that natural interest functions exist and indeed allow efficient maximization and hence automated ``discovery'' of dynamical phenomena which, until very recently, would have been unknown. The advantage of searching for something known is, of course, that we know what we are looking for, and the challenge lies in starting from as few assumptions as possible about where to look.
In summary, we provide here strong evidence that the proposed program of running quantum computers in ``discovery mode'' can work in principle.
Indeed, we anticipate that the design of interest functions will come to be a significant activity in many-body physics as quantum platforms begin to live up to their long-term promise. Likewise, we can imagine increasingly sophisticated approaches to search going forward. Together they suggest a vision of CERN-like facilities for many body physics combining state of the art quantum platforms with powerful classical computational infrastructure. 
Though the initial examples are, as necessitated by the limitations of currently available quantum devices, counterfactual validations of discovery mode they offer compelling evidence that the approach will be useful once quantum hardware can generate high-fidelity dynamics beyond classically simulable regimes.

At the point of writing, machine learning has been widely employed in quantum many-body physics, with early applications including phase classification and classical simulation \cite{carrasquilla2017ml_phases, vonNieuwenburg2017confusion,zhang2017ml,huang2022provably_efficient, carleo2017solving, miles2023discovery}, see, e.g., Ref. \onlinecite{carleo2019ml_review} for a comprehensive review.
In the context of quantum circuits, machine learning has been successfully employed for the design of pulse sequences for quantum gates as well as circuit synthesis for state preparation and error correction \cite{niu2019universal, floesel2018qec, metz2023rl, furrutter2024diffusion, faisal2024learning}, see Refs. \cite{gebhart2023review,krenn2023perspective} for recent reviews.
Our proposal is entirely different in spirit in that we propose to use machine learning (or other search algorithms) to systematically interact with and tune experimental setups and discover new \emph{physics}. Similar ideas have been developed in quantum optics in the form of MELVIN \cite{krenn2016automated, melnikov2018active}, which, while acting in an entirely virtual environment (and hence relying on the classical simulability of the proposed setups), still led to the development of new scientific ideas \cite{krenn2017entanglement,krenn2017quantum,gao2020computerinspired}. Moving beyond just physics, perhaps most notably, Ref. \onlinecite{szymanski2023autonamous} reported the operation of an automated laboratory for the discovery and synthesis of materials.
We note that in contrast to variational state preparation protocols, here we will be interested in properties of the unitary evolution itself, which, in contrast to specifying only the output when starting from a specific input state, does serve as the essential characterization of a physical system.

\section{Discovering Time Crystals: Interesting State Distributions}

\begin{figure*}
    \centering
    \includegraphics[width=1.0\textwidth]{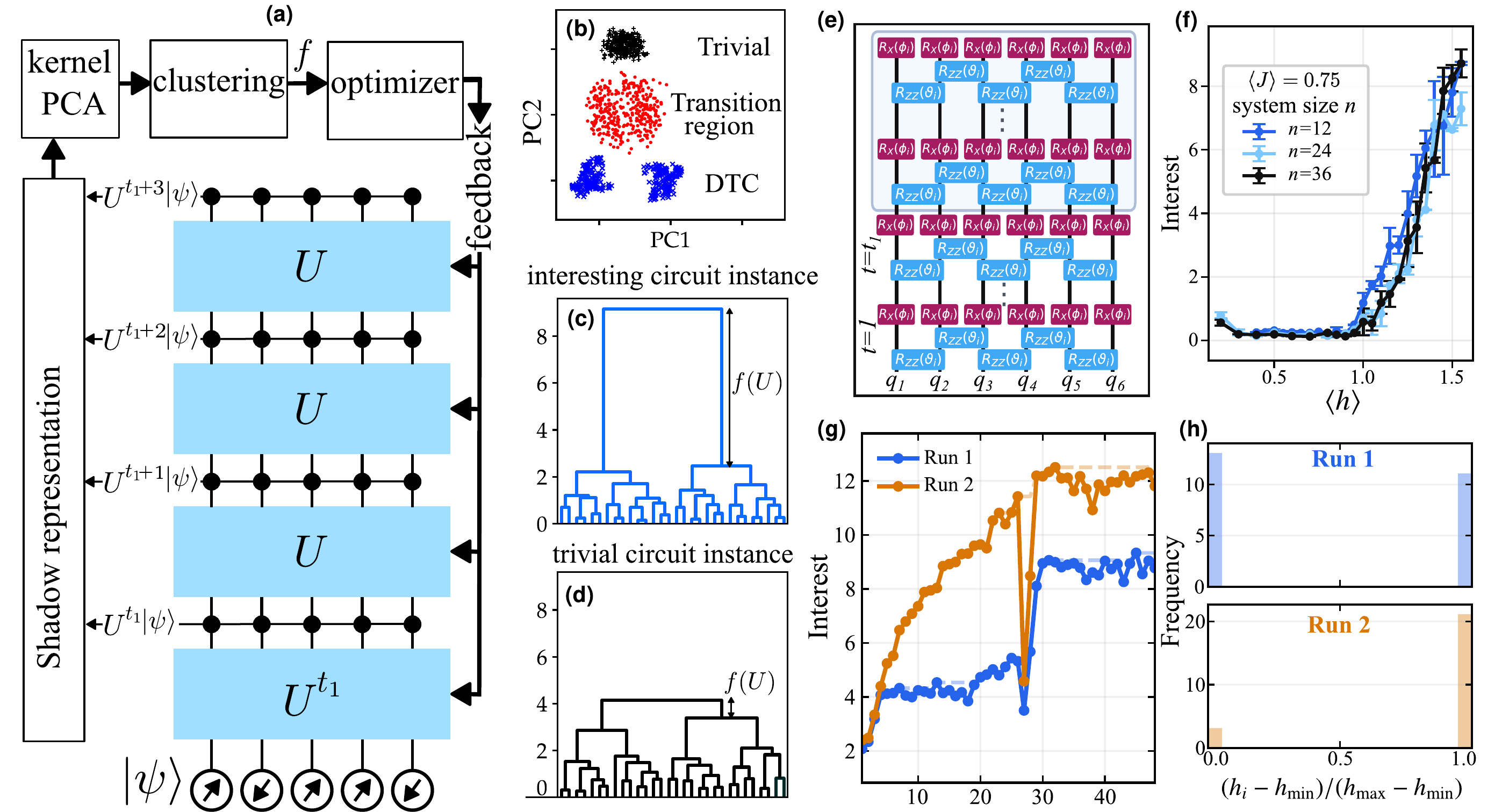}
    \caption{{\bf Discovering Time Crystals.} 
    One way to find interesting unitaries is to incentivize classifiability of the set of states $S(U,|\psi\rangle)=\{U^t |\psi\rangle \}_{t=t_1,\dots,t_1+T-1}$ produced by a sequential action of $U$ on a randomly picked computational basis state $|\psi\rangle$. (a) We use hierarchical agglomerative clustering in a low dimensional principal component (PC) space representation of the set $S(U,|\psi\rangle)$. Cluster separation is taken as the interest function $f(U)$ to be optimized.  (b) PC representation of states generated by representative examples of trivial and DTC unitary circuits.     (c,d) Dendrogram representation of the clustering, showing examples of an interesting and a non-interesting unitary. The height of the last leg of the dendrogram quantifying the difference between the last two Ward-linkage distances.
    (e) Kicked Ising type Floquet unitary circuit implemented on IBM Pittsburgh and IBM Boston quantum processors. States to be classified are picked from $T$ time steps after $t_1$th time step.
    (f) Interest function estimated in $12,24$ and $36$ qubit devices for kicked-Ising circuits as a function of site-averaged kick-strength $\langle h\rangle$ at fixed Ising coupling $J$ demonstrating classifiability of states as a robust indicator of the DTC phase. (g) Interest function vs iterations of representative COBYLA-based optimization trials executed with the help of a 24-qubit quantum processor. (h) Distribution of $h$-values in the optimized circuits generated by the two runs. Run 1 converged to a local optimum containing short regions of DTC phase interspersed by regions of vanishing $h$. Run 2 reached a better optimum with larger DTC regions producing better interest function.    \label{fig:discoverTimeCrystal}}
\end{figure*}

As our first example, we consider a class of (Floquet) unitaries involving the repeated action of a finite-depth unitary circuit $U$ which takes an initial state $|\psi\rangle$ to a sequence of states $S\equiv \{U^t |\psi\rangle\}_{t_1\leq t< t_1+T}$ in the interval $t_1\leq t < t_1+T$. The late time states in $S$, for generic $U$, undergo thermalization and have featureless few-body reduced density matrices independent of the specific subsystem, the time $t$, and the initial state $\ket{\psi}$.
A natural set of interesting, atypical circuits $U$ would be those which evade this fate, that is whose few-body observables have some nontrivial dependence on the quantities outlined above.  A simple way this could happen is by the state sequence having few-body features that fall into clusters.
As we show below, we can formulate the search for such unitaries as a maximization of an interest function that quantifies the classifiability of the states $S$ by an unsupervised learning algorithm (schematically shown in Fig.~\ref{fig:discoverTimeCrystal}(a)). We find, from extensive simulation of the optimization of parametrized unitaries, that the optimizer systematically and reliably discovers a discrete time crystal \cite{Khemani2016Phase,Else2016Floquet,RussomannoPRB17_LMGDTC, khemani2019brief,ElseReview20,Sacha17review,else2017prethermal,collura2022discrete}.

The states in $S$ have a dimension exponentially large in the number of qubits $n$ and therefore need to be represented in a convenient form to test classifiability. We use a classical shadow representation~\cite{Aaronson2020, Huang2020Predicting,Elben2018Renyi,elben2023review} of the states $S$, leveraging the ability of the quantum device to rapidly run the circuit multiple times to perform measurements. 
Each state is represented by a set $\{(s_1,s_2,\dots, s_n)_c\}_{c=1\dots N_{\rm s}}$ of $N_{\rm s}$ all-site Pauli measurements. Each element contains an ordered sequence of random Pauli measurement outcomes $s_i\in \{\pm1_x,\pm1_y,\pm1_z\}$ on the sites $1\leq i\leq n$  of an $n$ qubit-chain. 
The $N_{\rm s}n$ dimensional representation is still too large to run a classifier directly on this space, as the number $N_{\rm s}$ needed for a good shadow representation can be large; therefore we use a kernel principal component analysis (PCA) using the shadow kernel~\cite{huang2022provably_efficient,lewis2024improved} for dimension reduction. The kernel can be tuned to emphasize non-trivial structures in the features of the system's time series of few-body reduced density matrices to detect the breakdown of thermalization.
More precisely, the PCA is performed inside an extended feature space of all multi-site reduced density matrices and their powers but with a higher weight on few-site density matrices (see Methods). 

To look for signatures of non-thermalization in a series of states at times $t_1<t<t_1+T$, one  could investigate any atypicality in the distribution of these states in the principal component (PC) space (for instance as shown in Fig.~\ref{fig:discoverTimeCrystal}(b)). Atypicality in the form of clustering can be detected using hierarchical agglomerative clustering (HAC). HAC iteratively merges the nearest pair of clusters (based on Ward's linkage distance measure) among all  clusters of state-data, until a single one is eventually left; this can be visually represented by a dendrogram (Fig.~\ref{fig:discoverTimeCrystal}c,d). The distance between the last two merged clusters and their individual sizes quantifies the \emph{binary} classifiability of the dataset, and for the rest of this section we take this quantity as interest function $f(U)$.

As the estimated interest function has statistical fluctuations from the random choice of the initial state $|\psi\rangle$ and the quantum noise from Pauli measurements (in addition to the noise from the quantum device), gradient-free optimizers such as COBYLA and Nelder-Mead are better suited for optimization (see Methods). The statistical noise in the interest function can also be reduced by averaging it over $N_{\rm init}$ initial states $|\psi\rangle$ sampled from the computational basis. The estimation of the interest function thus involves $\sim N_s N_{\rm init} T t_{1}$  applications of the unitary,  which the quantum computer can efficiently perform. Figure \ref{fig:discoverTimeCrystal}(a) schematically shows the feedback loop between the quantum and classical devices to optimize the unitary.

\subsection*{Optimization trials using quantum processor}
To conclusively demonstrate our protocol,  we consider a parametrized unitary of the following form executed on a quantum processor
\begin{align}
&U(\{J_i\}_{i=1,\dots,n},\{h_i\}_{i=1,\dots n}) = e^{\imath H_J} e^{\imath H_h}\label{eq:parametricUnitary_DTC_1}\\
&H_J=\sum_{i=1}^{n-1} J_i Z_i Z_{i+1},\;\;H_h=\sum_{i=1}^n h_i X_i\nonumber
\end{align}
made of nearest-neighbor Ising interaction between qubits in one layer of the unitary and qubit rotations along a transverse axis in the second. 
Here, $X_i,Y_i,Z_i$ represent the Pauli matrices on $i^{\rm th}$ qubit. Figure~\ref{fig:discoverTimeCrystal}(f) shows the interest function evaluated on IBM Pittsburgh quantum processor with 12, 24 and 36 qubits and running Floquet unitary circuit of the form in Eq.~\ref{eq:parametricUnitary_DTC_1} as illustrated in Fig.~\ref{fig:discoverTimeCrystal}(e) parametrized by  $J_i\sim 0.75\pm \delta J_i$ and $h_i\sim \langle h\rangle + \delta h_i$ with $\delta J_i, \delta h_i$ picked from uniform distribution over [-0.2,0.2].
The interest function was averaged over two disorder realizations and four initial computational-basis states, using states from time steps $36$ to $43$, with each state represented by $192$ classical shadows.
The interest function faithfully captures the emergence of DTC order as $\langle h\rangle$ is tuned towards $\pi/2$ corresponding to the ideal DTC phase. 

Having demonstrated the ability of state classifiability to reflect interesting dynamics, we proceed to demonstrate the feasibility of interactively optimizing the circuit based on interest function values estimated using the quantum processor. 

Figure~\ref{fig:discoverTimeCrystal}(g) shows two 24-qubit COBYLA runs in which $\{h_i\}$ were optimized and the uniform coupling was restricted to $0.74\leq J\leq0.75$. Each interest-function evaluation was averaged over $N_{\rm init}=4$ initial computational-basis states with every evolved state reconstructed from $N_{\rm s}\sim 100$ classical shadows over $T=8$ Floquet time steps starting from time step $24$ for Pittsburgh device and $28$ for Boston.

In Fig.~\ref{fig:discoverTimeCrystal}(h), we show the distribution of $h_i$ values in the  circuits obtained at the end of two different optimization attempts. The two runs converged to qualitatively different optima. In Run 1 the optimizer settled on a strongly bimodal profile, with each $h_i$ pushed either to $h_i\approx\pi/2$ (the ideal DTC kick) or to its lower bound, so the chain breaks into several disconnected large-$h$ segments separated by sites with $h\sim 0$. Each large-$h$ block forms a DTC region with a period-doubled order with the low-$h$ sites forming their boundaries. Run 2, by contrast, converged to a configuration forming a large DTC region where $h_i\sim \pi/2$. 
The interest function is higher for Run 2, indicating that the classifiability metric rewards a large coherent DTC bulk. This is consistent with the expectation that the ideal DTC phase is expected only in infinite systems and small finite chains will have finite lifetime for the period two oscillations. 
Also, there is a tendency of $h_i$ to be 0 at the boundaries. It is surprising that such configurations form local minima in the interest function landscape - the interest function decreases with $h_i$ on these sites if $h_i$ is increased from $0$ towards the better minimum likely to be present near $h_i\sim \pi/2$. 
This is likely a finite $T$ effect - classifiability metric measured using states from a finite number of steps may reward small DTC patches, delimited by $h_i\sim 0$ sites even though they may have finite lifetime, more than one DTC patch with $h_i$ in that site far from $\pi/2$ which may nudge the oscillations to be incommensurate with the period two oscillations.

\begin{figure*}
    \centering
    \includegraphics[width=1.0\textwidth]{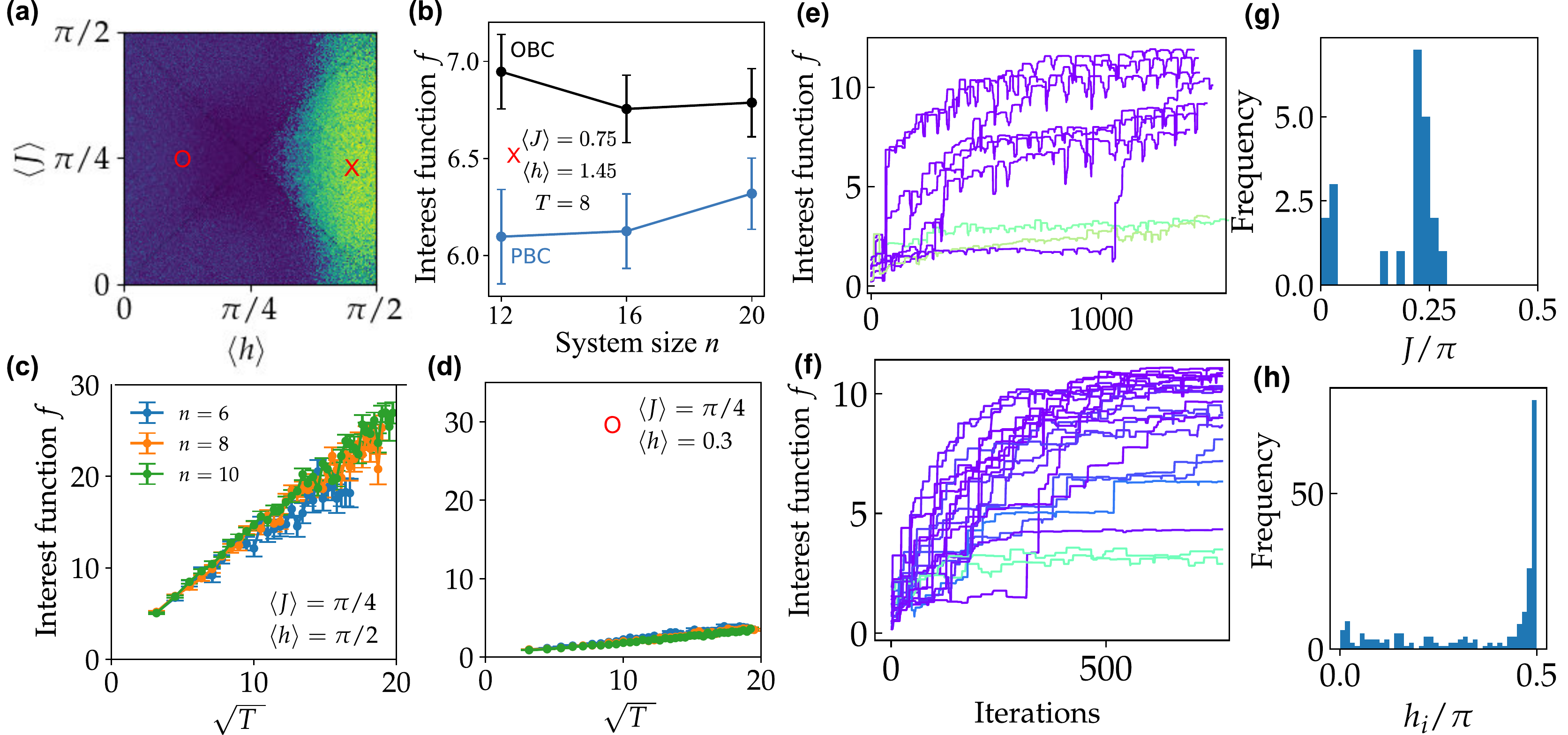}
    \caption{{\bf Classifiability metric.} 
    (a) Numerically obtained contour plot of disorder-averaged $f$ in the kicked-Ising model with disorder strength $\sigma_h=0.4$ and $n=6$ with periodic boundary conditions. The markers locate the circuit corresponding to the trivial (circle) and DTC phase (cross) state-set shown in Fig.~\ref{fig:discoverTimeCrystal}(b).  (b) Numerically obtained interest function in the DTC phase vs $n$ using simulations of a noisy circuit with $N_s=192$ shadows per state and interest function averaged over $N_{\rm init}=30$ random initial states, and using measurements on $T=8$ time steps from $t=36$ to $43$.
    (c) Disorder averaged interest function deep in the DTC phase vs the number of time steps $T$ over which state sequence data is collected. We find that the interest function scales as $f\sim n^0\sqrt{T}$.
    (d) Interest function in the trivial phase is much smaller and shows weak increase with $T$.
    (e,f) show the evolution of the interest function in numerical simulations of the optimization dynamics in a larger family of unitary circuits where the direction $\hat{s}$ of the  Ising interaction and the direction $\hat{m}$ of the transverse field are also optimized. Each curve represents an independent optimization run. Panel (e) shows results for a system with $n=6$ qubits keeping $J_i$'s also site-dependent. Different trajectories are colored based on the final value of $\hat{s}.\hat{m}$ with purple and green values showing cases where $\hat{s}.\hat{m}\approx 0$ (the optimal, kicked Ising regime) and $1$ (corresponding to an Ising model with longitudinal fields) respectively (f) shows the case where $J_i$ is taken to be site-independent. (g) Histogram of final $J$ for all optimization trajectories. (h) Corresponding distribution of final $h_i$.\label{fig:discoverTimeCrystal2}
    }
\end{figure*}

\subsection*{Classifiability-based interest function}
The classifiability-based interest function, as demonstrated above, provides a scalable way to detect deviations from thermalization. While it plays the role of an order parameter, it is structurally different from familiar order parameters measured in physics.

Conventional broken symmetry phases are characterized by a local order parameter which in the ordered phase scales (asymptotically) linearly with the size of the system in which they are measured and vanish exponentially with system size in the trivial phases. It is useful to ask how the interest function scales with the size of dynamical system in the system size $n$ and observation window $T$ in the light of measurements in large quantum processors across different system sizes shown in Fig.~\ref{fig:discoverTimeCrystal}(f).

In Fig.~\ref{fig:discoverTimeCrystal2}, we present further results using  numerical simulations to understand properties of the interest function. Figure~\ref{fig:discoverTimeCrystal2}(a) shows the contour plot of the interest function evaluated on 6-qubit circuits running the kicked Ising model with periodic boundary conditions parameterized by $\langle  J\rangle$ and $\langle  h\rangle$ ($J_i,h_i$ are picked from uniform distribution of width 0.2 about the mean). The interest function distinguishes the DTC phase from the Floquet paramagnetic and ferromagnetic phases producing a phase diagram expected in the kicked Ising model~\cite{Khemani2016Phase}. 
Interest function is intensive ($\sim n^0$) in the system size, at leading order, as revealed by the simulations shown in Fig.~\ref{fig:discoverTimeCrystal2}(b,c) and in quantum measurements shown in Fig.~\ref{fig:discoverTimeCrystal}(f) but it scales with the observation time window as $\sim T^{0.5}$ in the DTC phase. In contrast the interest function is much smaller and has weak system size and $T$ dependence in the trivial phase (Fig.~\ref{fig:discoverTimeCrystal2}(d), Fig.~\ref{fig:discoverTimeCrystal}(f)). Weak $n$-independence was obtained also in the presence of noise in time evolving block decimation (TEBD) simulations of the interest function in larger systems. Figure~\ref{fig:discoverTimeCrystal2}(b) shows the TEBD results as $n$ is varied from 12 to 20 in a noisy circuit with depolarisation probability $p_{\rm depol}=2\times10^{-3}$, amplitude-damping and dephasing probabilities $p_{\rm amp}=p_{\rm deph}=3\times10^{-4}$ in each qubit and after each brickwall layer of the kicked Ising circuit.

The weak system-size dependence of the interest function also suggests a continuation strategy for larger searches. One may first optimize a restricted circuit family on a small system, embed or tile the resulting parameters into a larger circuit, and subsequently fine-tune while gradually releasing additional degrees of freedom. The observation that the optimizer first learns global structural features, such as the relative interaction and field directions, before local coupling values (see App.~\ref{app:time_crystals}) is consistent with such a hierarchical search.

\subsection*{Optimizability of the interest function in more general circuits}
We now consider the question of whether the interest function can be used for optimization of a more general class of unitary circuits. For this we simulate the optimization dynamics using classical numerical calculations, and consider circuits with Floquet unitaries without a $\mathbb{Z}_2$ symmetry:
\begin{align}
&U(\{J_i\}_{i=1,\dots,n},\hat{s},\{h_i\}_{i=1,\dots n}, \hat{m}) = e^{\imath H_J} e^{\imath H_h}\label{eq:parametricUnitary_DTC}\\
&H_J=\sum_{i=1}^{n-1} J_i (\vec\sigma_i.\hat{s})(\vec\sigma_{i+1}.\hat{s}),\;\;H_h=\sum_{i=1}^n h_i \vec\sigma_i.\hat{m}\nonumber
\end{align}
made of nearest-neighbor Ising interaction between qubits in one layer of the unitary and qubit rotations in the second. 
Here, $\vec\sigma_i$ represents the Pauli matrices on $i^{\rm th}$ qubit, $\hat{s}$ and $\hat{m}$ denote the directions of the Ising interaction and qubit rotation axes respectively. We initialize the optimizable parameters with randomly chosen values ($J_i,h_i\in[0,\pi/2]$, $\hat{m},\hat{s}$ on the sphere). Pauli shadow measurements are performed along orthogonal directions different from the quantization axis of the computation basis states to reduce any bias in the protocol. We use a Nelder-Mead gradient-free optimizer again to steer the parameters towards larger values of the interest function $f$.

Fig.~\ref{fig:discoverTimeCrystal2}(e,f) shows the interest function's evolution with iterations for an $n=6$ qubit system. (Further details of the optimization attempts can be found in App.~\ref{app:time_crystals}). In panels (e) and (f), we present, respectively, the results for the case where $J_i$s are set to a position-dependent, and where the $J_i$ are site-independent.

The optimized configuration is found to have $J\approx\pi/4$ with a high probability (Fig.~\ref{fig:discoverTimeCrystal2}(g)). The directions $\hat{s}$ and $\hat{m}$ of the Ising and the qubit rotation axis in the optimized configurations are found to be orthogonal to each other, implying recovery of the $\mathbb{Z}_2$ symmetry at the optimum. The distribution of $h_i$ values in the final configurations is shown in Fig.~\ref{fig:discoverTimeCrystal2}(h).

The distribution of $h_i$ has two peaks - a dominant one around $\pi/2$ and the small second one near $h_i=0$. The optimum is likely to be a finite-width distribution located around $\pi/2$.
As discussed above, small peak near $h_i\sim 0$  likely arises due to the optimizer converging to a local minimum in the cost function. Consistent with this, we find that increasing the learning rate (i.e. the simplex size of the Nelder-Mead optimizer) leads to a decrease in the fraction of sites with $h_i\approx 0$~(see App.~\ref{app:time_crystals}).

These optimal parameters can be seen to be those of a unitary in a DTC phase with a $\mathbb{Z}_2$ symmetry. Thus, an attempt at maximizing the binary classifiability of the state sequence generated by the quantum machine would lead us to a DTC unitary with $\mathbb{Z}_2$ symmetry. 

Our results in the quantum device and in simulations show that our protocol finds with high probability a close-to-``optimal'' DTC realization---as opposed to a random realization within the DTC phase. Moreover, consistent convergence into the period-2 DTC phase (as opposed to other solutions such as period-$4$ phase) is indicative of a generic feature that the optimizer is more likely to detect robust solutions rather than fine-tuned ones~\cite{keyserlingk2016absolute}. These results leave us with a subtle open question on the relation between the interest function and the order parameter of the discovered phase, as well as on the identification of the attraction basin around the optimum of the interest landscape with the corresponding one defined by more standard condensed-matter-physics-inspired metrics~\cite{khemani2019brief}.

The utility of state-distribution-based interest functions for discovering atypical non-thermalizing dynamics does not appear to be specific to time-crystalline order: in separate ongoing work, a related protocol has also counterfactually rediscovered many-body localization.

\section{Discovering Dual-Unitary circuits: Interesting Spectral Statistics\label{sec:discover_dual_unitaries}}

\begin{figure*}
    \centering
    \includegraphics{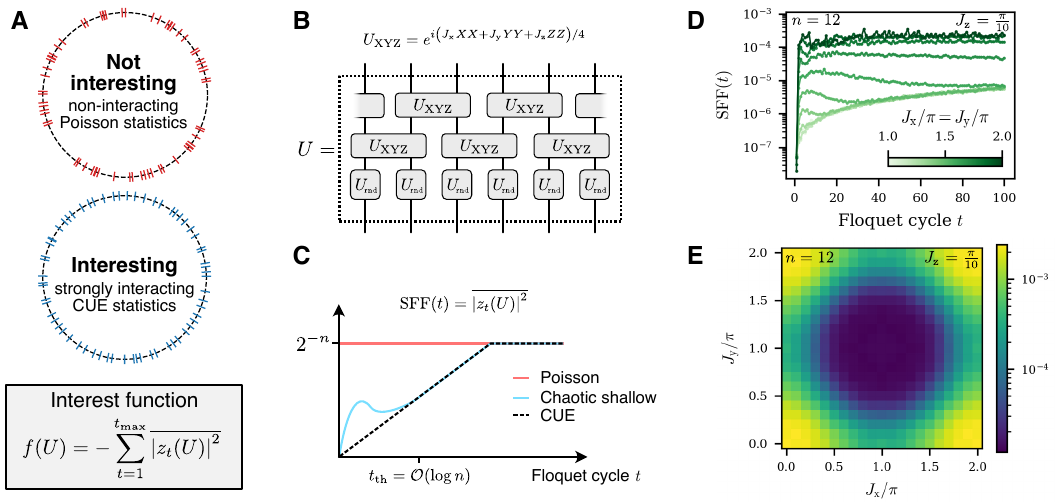}
    \caption{{\bf Discovering interesting unitaries.} In general, interest functions are functions $f: U(2^n) \to \mathbb{R}$. The simplest such functions are traces of integer powers $t$ of $U$, denoted here as $z_t(U)$ (Eq.~\ref{eq:ztU}), These quantities represent the moments of the eigenvalue distribution of $U$ on the unit circle, which behave very differently in systems of non-interacting qubits on the one hand, and in global fully random unitaries drawn from the so-called Circular Unitary Ensemble (CUE) on the other hand (A).  They are sensitive probes of spectral correlations, which may be used to identify "interesting" unitary circuits exhibiting extremal quantum chaos (possibly under certain global constraints). In a generic brickwork circuit family such as that illustrated in (B), traces of powers of $U$ have a characteristic  intermediate behavior between non-interacting and fully random unitaries (C). Maximizing $f(U)$ steers the parametrized finite-depth circuit towards the fully random unitary (CUE) behavior. Remarkably, $f(U)$ does attain the ``maximally chaotic'' value, despite the finite-depth constraint; its optima coincide with dual-unitary circuits. Our family of circuits interpolates between Poisson, generic, and dual-unitary behavior (D). Computation of the interest landscape showing $-f(U)$ demonstrating that an optimization cycle would successfully steer the parametrized circuit towards a dual-unitary submanifold in parameter space (E).
    }
    \label{fig:discover_dual_unitaries}
\end{figure*}

In the previous example we characterized the ``interest'' of a given unitary $U$ by classifiable patterns produced by its repeated applications to a product state. One might ask whether it is possible to characterize interesting properties of the unitary itself more directly, without reference to particular states.

When asking about basis-independent functions on the unitary group $U(2^n)\to\mathbb{C}$, one is naturally led to consider expressions of the form
\begin{equation}\label{eq:ztU}
    z_t(U) = 
    \frac{\tr U^t}{\tr \mathbb{1}}
    = 
    \frac 1 D \sum_{j=1}^{D} e^{i t\theta_j}
\end{equation}
where $D=2^n$ and $e^{i\theta_j}$ are the eigenvalues of $U$. For $t=1$, this function---i.e., the normalized trace of $U$---characterizes the center of mass of the distribution of eigenvalues, and for larger integer values of~$t$, it characterizes the higher moments of this distribution, i.e., $z_t(U)=\int_{0}^{2\pi} d\theta \, \rho(\theta) e^{it \theta}$, with $\rho(\theta)=\frac 1D\sum_{j=1}^D \delta(\theta-\theta_j)$.
This quantity is related to the so-called spectral form factor (SFF) $\abs{z_t(U)}^2= \int_{0}^{2\pi}d\theta \int_{0}^{2\pi}d\theta' \, \rho(\theta)\rho(\theta') e^{i t(\theta-\theta')} $, which plays a pivotal role in quantum chaos theory \cite{berry1985semiclassical}.

For any~$t$,  the absolute value $|z_t(U)|$ can be maximized by trivial unitaries such as the identity, which yields $z_t(U)=1$, as well as by simple product unitaries, acting non-trivially on individual qubits (or few qubits).
More generally, if $U$ can be broken down into a tensor product $\bigotimes_{i=1}^n U_i$ of $n$ $2\times 2$ unitaries, where $U_i$ has eigenvalues $e^{i(\alpha_i\pm \phi_i)}$, then $|z_t(U)|^2= \prod_{i=1}^n \cos^2 (\phi_i t)$. For such ``non-interacting'' unitary circuits, maximization is realized when each $U_i$ is a rotation by an angle multiple of $\pi/t$ around arbitrary axes.

\emph{Minimization} of $|z_t(U)|$ represents, instead, a far less trivial problem for unitary circuits.
In fact, for non-interacting unitaries as above,  randomly chosen eigenphases $\alpha_i\pm\phi_i$ lead to values of $z_t(U)$ distributed around the origin in the complex plane  with broad\footnote{in fact, log-normal} absolute value fluctuations of order $|z_t(U)|^2 \approx 1/D$ for all $t\ge1$.
While it is often easy to construct examples of non-interacting unitaries with fine-tuned relative eigenphases $\phi_i$ such that $z_t(U)=0$ for a pre-defined set of values of $t$,\footnote{As an extreme example, a spectrum formed by the set of $2^n$-th roots of the identity gives $z_t(U)=0$ for all $t$, except for integer multiples of $2^n$; it can be checked that the choice $\phi_i=\pi \, 2^{i-1-n}$ for $i=1,\dots,n$ produces this spectrum.} arbitrarily small perturbations restore values $|z_t(U)|^2\approx 1/D$. 
On the other hand, it is a well-known fact~\cite{Diaconis1994On, Mehta2004Random} that fully random unitaries in $U(D)$ have strong statistical correlations between eigenvalues, formally equivalent to a gas of charged particles constrained to the unit circle and experiencing Coulomb repulsion. In this case, $z_t(U)$ has much smaller fluctuations around the origin in the complex plane, of order $|z_t(U)|^2\approx t/D^2$, as long as $ t <D$. Contrary to the non-interacting examples above, such a smallness is---by construction---robust to perturbations. The search for ensembles of unitaries with suppressed fluctuations of $z_t(U)$ is thus related to the discovery of strongly  correlated spectra, characteristic of extremal quantum chaotic behavior. This dichotomy is illustrated in \autoref{fig:discover_dual_unitaries}A.

Generic finite-depth unitary circuits exhibit, in a sense, behavior intermediate between non-interacting and fully random unitaries, with absolute value fluctuations of order $|z_t(U)|^2\approx 1/D$ for $ t \ll t_{\rm Th}(n)$, and  of order $|z_t(U)|^2\approx t/D^2$ for $ t_{\rm Th}(n) \ll t <D $, where the scale $t_{\rm Th}\propto \log n$, known as Thouless time, grows unbounded with system size~\cite{kos2018many,chan2018spectral}, see \autoref{fig:discover_dual_unitaries}B,C.
Simultaneous minimization of $|z_t(U)|$ for all $t$ may thus be conjectured to yield ``interesting'' ensembles of finite-depth unitary circuits with anomalously strong spectral correlations over all scales, similar to fully random unitaries. It is natural to ask whether such circuits exist, what their physical properties are, and how easy it would be to discover them.
We will show that running quantum computers in discovery mode without prior knowledge,  this line of reasoning could have led one to anticipate the discovery of a class of circuits known as dual-unitary, analytically identified by Bertini, Kos and Prosen in 2019~\cite{bertini2019exact}, which exhibit extremal spectral correlations and spread quantum information at the fastest possible speed~\cite{bertini2021sff, Bertini2019Entanglement, Zhou2022Maximal, bertini2025review}.
More generally, our approach suggests a straightforward route to future discoveries, by means of interest functions constructed as physics-inspired generalized functionals of~$z_t(U)$.

We begin by writing down a general interest function
\begin{equation}
\label{eq:fU_dual_unitary}
    f(U) = -\sum_{t=1}^{t_{\rm max}} V_t \big[z_t(U)\big]
\end{equation}
where the ``potential'' $V_t$ takes the form
\begin{equation}
\label{eq:fU_landau}
    V_t(z) = \frac{\alpha_{1,t}}2 |z|^2 + \frac{\alpha_{2,t}}4 |z|^4 + \dots
\end{equation}
The specific problem discussed above, i.e., the simultaneous minimization of $|z_t(U)|$ for several values of $t$ over a fairly generic parametrized finite-depth circuit, corresponds to choosing non-negative coefficients $\alpha_{n,t}\ge 0$ in Eq.~\eqref{eq:fU_landau}.
The choice $\alpha_{1,t}=2$, $\alpha_{n>1,t}=0$, yields, in particular, the (negative) time-integrated  SFF. 

We consider here the family of  circuits with architecture and parametrization shown in \autoref{fig:discover_dual_unitaries}B, with a layer of single-qubit gates followed by a brickwork layer of symmetric two-qubit gates, parametrized as $U_{\rm XYZ} = \exp(\tfrac{i}{4} (J_{\rm x} X  X + J_{\rm y}  Y Y + J_{\rm z} Z  Z))$. By suitably adjusting the parameters, this circuit family covers all possible brickwork circuits. For the purpose of this work, we evaluate the interest function on a classical computer for small systems with $n\le 14$. Since the relative width of circuit-to-circuit fluctuations of $z_t(U)$ does not shrink with increasing system size~\cite{prange1997the}, it is challenging to resolve the variation of $f(U)$ and perform an optimization.

We have thus upgraded $f(U)$ to an average over an ensemble of circuits with independent and identically distributed Haar-random single-qubit gates, denoted $\overline{\cdot(U)}$.  Due to disorder averaging, simulating a full adaptive optimization on many parameters, as was done for the DTC in the previous section, is computationally expensive. However, we show in \autoref{fig:discover_dual_unitaries}D and E that the interest landscape is locally concave and attains its maximum at the parameter submanifolds $J_{\rm a} = J_{\rm b} = \pi$ with $a,b=x,y,z$, $a\neq b$, corresponding to the dual-unitarity condition. This maximum value coincides with the theoretical result attained by fully random unitaries. 

Additional insight on the structure of the landscape of $f(U)$  can be obtained from recent results in the theory of quantum many-body chaos, which predict (assuming periodic boundary conditions)~\cite{chan2018spectral,garratt2021local,yoshimura2025dynamics}
\begin{equation}
\label{eq:prediction}
    \overline{|z_t(U)|^2} = \frac{t}{D^2} \left[ 1 + \binom{n}{2}(t-1)e^{-2t/\tau}  + \dots \right].
\end{equation}
In this equation, the parameter $\tau=\tau(U)\ge 0$---interpretable as an inverse ``domain-wall tension'' for effective pairing degrees of freedom in spacetime---controls the Thouless time $t_{\rm Th}(n)$. Validity of the universal random-matrix-theory prediction at arbitrarily short times, realized by dual-unitary circuits, requires $\tau = 0$. Assuming $\tau$ to increase smoothly as $U$ is detuned away from the dual-unitary parameter submanifold, the interest function's variation around the maximum is
\begin{equation}\label{eq:fU_exact}
    \delta f(U) \approx  - n^2 e^{-4/\tau(U)\, }    \, .
\end{equation}
This is a locally concave landscape around the maximum at $\tau = 0$. The form of \autoref{eq:fU_exact} suggests that the maximum is very broad, with all derivatives vanishing at the maximum. This hypothesis is consistent with the numerical data in \autoref{fig:discover_dual_unitaries}E, plotted on a logarithmic scale. This feature is interesting in its own regard, as it suggests a somewhat surprising stability of spectral signatures of dual-unitarity to perturbations.
Further analysis of the prediction~\eqref{eq:prediction} suggests that at fixed distance from dual-unitarity (hence, fixed $\tau$), the gradient increases exponentially with system size. Nevertheless, we note that even for the  limited system sizes considered here, the gradient only becomes unresolvable quite close to dual-unitarity (see App~\ref{App:DualU}).

As discussed in~\appref{app:SFF}, experimental estimation of the SFF is possible in principle, but challenging for large systems, due to an unfavorable sample complexity.

A closely related quantity, which is more efficiently accessible with quantum devices~
\cite{Collins2023sff_experiment,dong2025sff_experiment}, is the \emph{partial} spectral form factor (pSFF)~\cite{joshi2022pSFF}. As we discuss in detail in~\appref{app:SFF}, this simpler quantity is also a faithful witness of extremal quantum many-body chaos, and, in particular, it takes lower values for dual-unitary circuits than for generic finite-depth unitary circuits \cite{fritzsch2024pSFF,yoshimura2025dynamics}.
We therefore expect minimization of the pSFF to be an excellent proxy for finding maximally chaotic (and non-generic) circuits as well, and hence to give rise to an alternative, potentially more efficient discovery protocol. This distinction illustrates a general requirement of discovery mode: access to large quantum dynamics must be paired with interest functions whose relevant variations can be resolved with feasible sample complexity.

We finally note that the form of our interest function in~Eq.~\eqref{eq:fU_dual_unitary} is suggestive of further physics-inspired discoveries. For example, consider the choice $\alpha_{1,t}<0$, $\alpha_{2,t}>0$, $\alpha_{n>2,t}=0$, corresponding to a ($t$-dependent) Mexican-hat potential. Maximization of this interest function will steer the circuit to realizations where $z_t(U)$ exhibits the weakest possible fluctuations around a circle of finite radius $\sqrt{|\alpha_{1,t}|/\alpha_{2,t}}$ in the complex plane. It is straightforward to see that this property is associated with a smoothly modulated density of eigenvalues $\rho(\theta)=\frac{1}{2\pi} \big( 1 + 2 \sum_{\ell=1}^\infty c_{\ell} \cos(\ell\theta) \big)\ge0$ on the unit circle, described by Fourier harmonics $c_{\ell>0}=\sqrt{|\alpha_{1,\ell}|/\alpha_{2,\ell}}$ directly related to the Mexican-hat potential.
Such a smooth spectral density modulation on top of a strong level repulsion is suggestive of invariant unitary random matrix ensembles~\cite{grosswitten,Mehta2004Random}.
To the best of our knowledge, classes of finite-depth circuits possessing such novel ``interesting'' spectra are yet to be discovered.  

\section{Conclusion and Outlook}

In this work we have proposed and demonstrated a novel functioning mode of quantum-computing platforms in conjunction with classical optimization and learning techniques, which we called ``discovery mode". 

As summarized in Fig.~\ref{fig:toc}, this approach hinges on the design of ``interest functions'', i.e., functions sensitive to conjectured features in data collected from parametrized quantum dynamics.

Classical optimization of the interest function over the parameters drives the search through a space of experimentally realizable quantum evolutions, with the aim of identifying concrete new and possibly unknown unitary instances exhibiting features abstractly defined in the interest function.

The inverse-problem nature of this approach and the heuristic aspect of the search for conjectured features represent the main divide between discovery-mode quantum experiments and existing hybrid quantum-classical variational techniques~\cite{cerezo2021variation_review, Bharti2022Noisy}, which target, for example, the ground state of a specific Hamiltonian. 
In common with previous hybrid schemes, quantum computing resources provide input data for classical optimization, which feeds updated parameters back into the quantum device in an iterative learning loop.

The examples studied in this work provide evidence that an unbiased use of quantum computers in discovery mode can lead to important breakthroughs in quantum many-body dynamics without explicit prior knowledge of their microscopic realization. In particular, we demonstrated the possibility of the discovery of genuine non-equilibrium phases of matter (such as DTCs) and of maximally chaotic quantum evolutions (dual-unitary circuits). The counterfactual discovery routes presented here are completely different from the historical discovery routes, found in 2016 and 2019, respectively - the examples here rediscovered the physics with approaches that had minimal encoding of the outcome. We were able to demonstrate the optimization scheme for the DTC example on a superconducting quantum processor with up to 36 qubits showing the feasibility of the protocol in real noisy devices.

Both examples naturally suggest ways forward, including alternative measures of deviation from thermalization, geometric probes of state manifolds in embedding spaces, and more general Landau-functional constructions. Even local optima of an interest landscape may reveal unexpected physics. Ongoing work provides two further indications of the breadth of the approach: a related many-body search has counterfactually rediscovered many-body localization \footnote{G.~J.~Sreejith and collaborator, unpublished.}, while a separate programme is finding transversal logical gates for quantum codes using searches that run effectively with modest classical resources \footnote{Work in progress; authors and citation to be added.}. The latter example emphasizes that the classical optimization loop need not itself be the fundamental bottleneck; quantum hardware becomes indispensable when evaluating the interest function requires access to dynamics beyond classical simulation.

More broadly, discovery mode separates two computational tasks. Quantum processors generate samples from many-body dynamics that may be inaccessible to classical simulation, while classical learning systems search over experimentally controllable parameters. The latter search need not be repeated independently from random initialization at every scale: it can be organized as a curriculum in system size, circuit depth, and parameter resolution, with learned solutions transferred from simpler instances to more complex ones. The central design problem is therefore to construct interest functions whose measurement cost and optimization signal remain controlled as the quantum system grows. With increasing scale and fidelity of quantum platforms, discovery mode provides a new framework in which they can aid scientific discovery.

\vspace{1.5em}
\centerline{\bf Acknowledgements.} 
\vspace{0.5em}
We thank John Chalker, Pieter Claeys, Andrew Daley for helpful discussions. We also thank Stefano Veroni for his involvement in the early stages of the project and for useful discussions, as well as Marcello Dalmonte, Vedika Khemani, and Siddharth Parameswaran for helpful comments on an early version of the manuscript.
We also thank Sebastian Brandhofer for helpful discussions about the implementation of the work on IBM devices.
B.P., A.L., and S.L.S. acknowledge support by the Leverhulme Trust via a Leverhulme Trust International Professorship to S.L.S (Grant Number LIP-202-014).
B.P. acknowledges support by the Humboldt Foundation via a Feodor Lynen Fellowship.
A.D. acknowledges support from the EPSRC through the QCi3 Hub (EP/Z53318X/1).
We thank NSM Param Bramha computing facility at IISER Pune for providing the computing resources used for part of the calculations presented here. G.J.S. acknowledges support from IHUB Quantum Technology Foundation and thanks the Rudolf Peierls Centre for Theoretical Physics, University of Oxford for their hospitality during the completion of this work.
\bibliography{references}

\clearpage
\vfil \break
\appendix

\section{Details of the classifiability-based interest function}
\paragraph*{Kernel PCA:} For an initial computational basis state $\vert \psi \rangle$, the time evolution by $U$ from $\vert \psi \rangle$ generates a set of states $S={U^t\vert \psi \rangle}_{t=t_1\leq t< t_1+T}$ between time steps $t_1$ and $t_1+T$. We represent each state $\vert \psi_t\rangle = U^t\vert \psi \rangle$  by its classical shadows represented in terms of projective Pauli measurement outcomes of all sites along one of the three orthogonal directions chosen independently randomly on the sites. Thus each state $\vert \psi_t\rangle$ is represented by collection  $\{(s^c_1,s^c_2,\dots, s^c_n)\}_{c=1,2,\dots N_s}$ of $N_s$ ordered sequences each containing the measurement outcomes $s^c_i\in \{\pm 1_x, \pm 1_y, \pm 1_z\}$ on sites $i=1,2\dots n$.
The kernel principal component analysis used for dimension reduction of the state data $S$ encoded using the shadows is defined by the kernel function which is an inner-product between the embedding of the states $|\psi\rangle\equiv \{(s^c_1,s^c_2,\dots, s^c_n)\}_{c=1,2,\dots N_s}$ and $|\underline\psi\rangle\equiv \{(\underline{s}^c_1,\underline{s}^c_2,\dots, \underline{s}^c_n)\}_{c=1,2,\dots N_s}$ in an extended space.

A convenient extended feature space~\cite{huang2022provably_efficient} that can be used here, contains the direct sum of all vectorized (i.e. flattened) $r$-site reduced density matrices (for all possible choices of the $r$ sites), and their $d$th tensor product powers (for $d=0,1,\dots$). An embedding of a state $|\psi\rangle$ in this extended feature space can be written in terms of its projective measurements $\{(s^c_1,s^c_2,\dots, s^c_n)\}_{c=1,2,\dots N_s}$ as:
\begin{equation}
\phi\left[\psi\right]=\bigoplus_{d=0}^{\infty}\sqrt{\frac{\tau^{d}}{d!}}\left(\bigoplus_{r=0}^{\infty}\sqrt{\frac{1}{r!}\left(\frac{\gamma}{n}\right)^{r}}\Omega_{r}\right)^{\otimes d}
\end{equation}
where $\Omega_r$ is the direct sum of the vectorized reduced density matrices on possible choices of $r$-sites:
\begin{equation}
\Omega_{r}=\oplus_{i_{1}=1}^{n}\dots\oplus_{i_{r}=1}^{n}\vec{\rho}_{i_{1}i_{2}\dots i_{r}}\nonumber
\end{equation}
with 
\[
\vec{\rho}_{i_{1}i_{2}\dots i_{r}}={\rm vec}\left[\frac{1}{N_s}\sum_{c=1}^{N_s}\otimes_{l=1}^{r}M^{-1}\left(s_{i_{l}}^{c}\right)\right],
\]
where $M^{-1}(a)=3\vert a \rangle \langle a \vert-\mathbb{1}_2$ is the inverse depolarization channel on a qubit. The hyperparameters $\tau$ and $\gamma$ parametrize the embedding and determine how much weight is put on features present in the $d$th tensor product power of the $r$-site reduced density matrices, respectively.

Principal component estimation relies only the Euclidean inner product between these extended space data points. This inner product between embeddings of two states $\psi$ and $\underline{\psi}$ is called the shadow kernel function 
\begin{equation}
K(\psi,\underline{\psi}) =\exp\left[\frac{\tau}{N_{s}^{2}}\sum_{c,\underline{c}=1}^{N_{s}}\exp\left(\frac{\gamma}{n}\sum_{i=1}^{n}k\left(s_{i}^{c},\underline{s}_{i}^{\underline{c}}\right)\right)\right]
\end{equation}
where $k(a,b)={\rm{Tr}}[M^{-1}(a)^\dagger M^{-1}(b)]$. We choose the hyperparameters $\tau$ and $\gamma$ to be $4$ and $0.1$ respectively. For small $\gamma$, the kernel focuses on properties only of few-site reduced density matrices. A more optimal choice of hyperparameters, and more generally the kernel function may lead to better performance of the optimizer. However we have not explored the hyperparameter dependence of the optimizer performance.

\paragraph*{Clustering:} Clustering was performed on the state set where each state was represented by its leading principal component. The HAC algorithm was used with the Ward linkage measure for the distance, with the Euclidean distance used in estimation of variances in the linkage~\cite{mullner2011modern}. The calculations presented here use the following conventions for HAC algorithm implemeneted by SciPy.

Firstly we define a linkage criterion $W$ that is used later to determine whether cluster $A$ and $B$ need to be merged.
\[
W(A,B)=\sqrt{\frac{2n_A n_B}{n_A+n_B}} \lVert\bar{x}_A-\bar{x}_B\rVert
\]
where $\lVert \cdot \rVert$ is the Euclidean norm and $\bar{x}_A$ is the centroid of $A$.
Starting from each data point (here the PC representation of each state)  forming its on single-element cluster, the algorithm, in each iteration, merges the pair of clusters with minimum $W$ into a single cluster. Thus the number of clusters reduces by one each iteration until only a single cluster is left. $W$ associates a cost to merging $A$ and $B$; with $W$ being smaller if the union of $A,B$ forms a compact cluster. 
A heuristic measure of binary classifiability is the difference between the linkage criteria in the last two steps \begin{equation}f=W_{T-1}-W_{T-2}.\end{equation} A large $W_{T-1}$ implies the last merge forced a merger of distant clusters and small $W_{T-2}$ implies that the merger in the penultimate step created a compact cluster.

\section{Additional data from measurements on the quantum device}

\begin{figure*}
\includegraphics[width=1\textwidth]{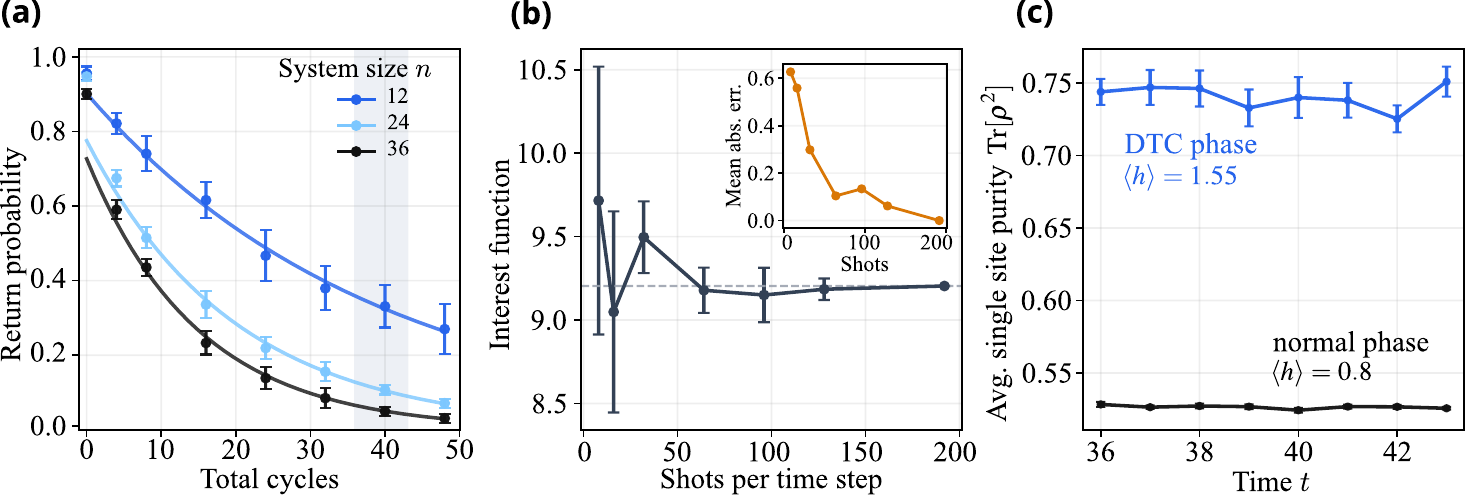}
\caption{ (a) Loschmidt echo as a function of the total number of forward and backward circuit cycles, characterizing the effect of circuit imperfections. Measurements were performed in a kicked Ising circuit with $J_i\sim 0.75\pm 0.2$ and $h_i\sim 1.3\pm 0.2$.
(b) Interest function in a DTC circuit as a function of the number of Pauli shadow measurements on the $24$-qubit quantum device (c) Average single site purity as a function of time step for $24$-qubit DTC and normal state circuit. 
\label{fig:device_returnprob}}
\end{figure*}

The protocol described in the main text for the optimization of the classifiability-based interest function was run on the IBM Pittsburgh superconducting quantum processor, using fractional/native two-qubit gates. 

Figure~\ref{fig:device_returnprob} collects three independent checks that validate the data obtained on hardware: (a) a mirror-circuit benchmark that quantifies how far the device can be evolved before noise dominates, (b) a demonstration that the shadow estimate of the interest function self-averages with the number of measurement shots, and (c) the average single-site purity, which serves as an independent, model-free diagnostic that separates the two dynamical phases.

In the mirror-circuit benchmark [Fig.~\ref{fig:device_returnprob}(A)], we estimate how deep the Floquet dynamics can be probed before device noise strongly suppresses a global coherence diagnostic. We run a mirror (Loschmidt echo) circuit: the system is evolved, starting from a computational basis state, for a number of Floquet cycles and then evolved by the exact inverse sequence. An ideal noiseless device returns to the initial state with unit probability. The measured return probability decays approximately exponentially with the total number of applied cycles. As expected, larger systems decay faster because more entangling gates are applied per cycle. The shaded region marks circuit depths comparable to those used for the late-time state collection. Although the all-bit return probability is already substantially suppressed in this regime, it remains above the noise floor, indicating that the device has not completely lost dynamical information.
This benchmark should be interpreted as a strict global diagnostic since a single bit flip can remove a shot from the return-probability signal. It is therefore not a direct proxy for the information retained in local observables or in few-site reduced-density-matrix features.

Figure~\ref{fig:device_returnprob}(b) shows the interest function calculated for a kicked Ising circuit in the DTC phase as a function of the number of shadows (using randomized single-qubit Pauli measurements). A few hundred shots per time step are sufficient for a stable estimate of $f(U)$. This is what makes the protocol practical on present-day hardware, even where the shot budget is limited. 

Finally, in Fig.~\ref{fig:device_returnprob}(c), we demonstrate an additional cross-check on the circuit and measurements on the quantum device. We reconstruct the average single-site purity $\mathrm{Tr}(\rho_i^2)$ from the collected shadow data. In the DTC phase the spins remain localized and weakly entangled with their environment, so the single-site reduced density matrices stay close to pure. In the thermalizing (normal) phase the reduced density matrices relax towards the maximally mixed state and the purity approaches $1/2$. The two phases are cleanly separated and the separation is stable over the entire measured time window, confirming that the hardware and the shadow measurements faithfully distinguish the dynamical regimes that the discovery protocol works on.

\section{Supplemental details for the optimizer for state classifiability\label{app:time_crystals}}
\begin{figure*}
\includegraphics[width=1\textwidth]{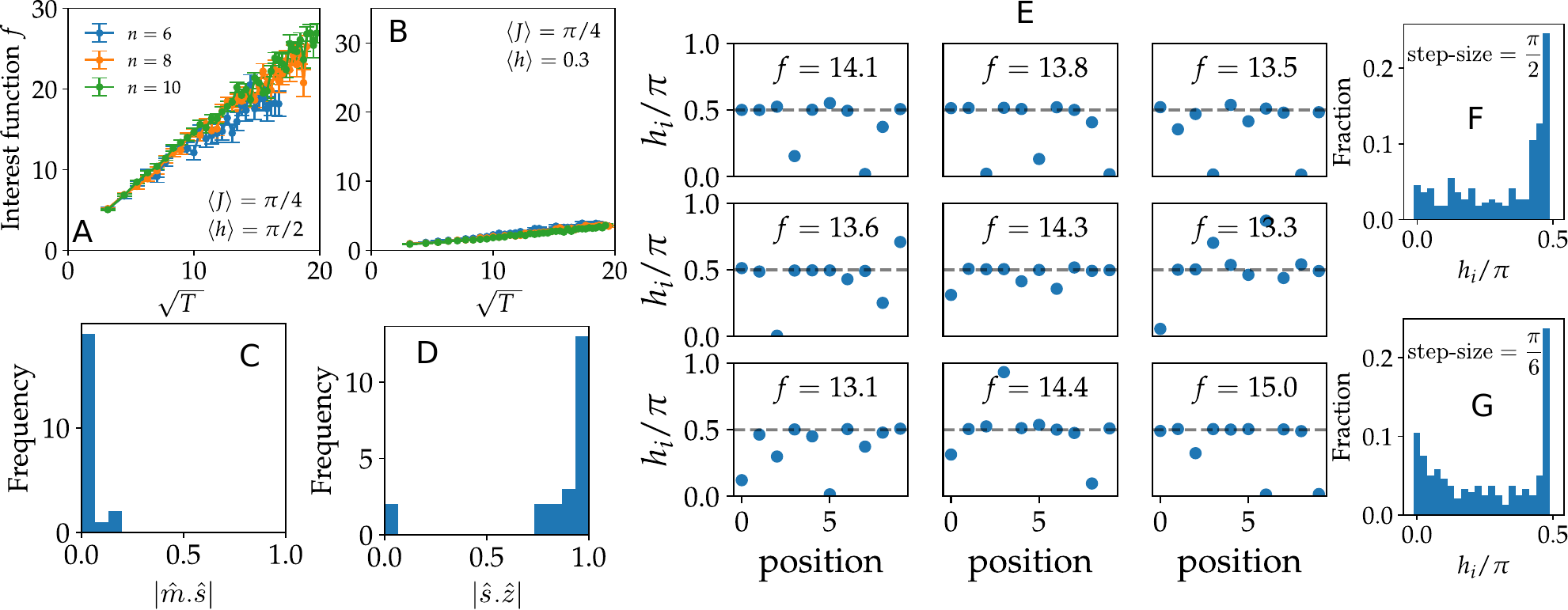}
\caption{
(A,B) Dependence of the interest function $f(U)$ on the number of time steps $T$ over which the data was collected. Panels (A),(B) show this for unitaries deep in the DTC phase and trivial phase respectively. We empirically find that $f$ scales as $T^{1/2}$ and is independent of the number of qubits $n$ in the DTC phase. (C) Distribution of the angle between the Ising interaction direction $\hat{s}$ and the direction of the field $\hat{m}$ in the final configurations given by the optimizer. The overlap should be 0 for the unitary circuit to have $Z_2$ symmetry. (D) Distribution of the angle-cosines between the Ising interaction direction ($\hat{s}$) and the quantization axis $\hat{z}$ of the qubits. (E) Sample spatial profiles of the field $h_i$ at the end of optimization runs, shown for the 9 optimization runs with largest $f$. (F,G) Distribution of the field values in the configurations generated by the optimizer at the end of 500 iterations. Panels (F) and (G): show the results from optimization with initial Nelder-Mead simplex sizes  $\pi/2$ and $\pi/6$ respectively. The results presented in panels C-G are generated with $n=10$ qubits and hyperparameters $N_s=500, N_{\rm init}=32, t_1=10$ (initial time step from where states are collected to test classification), and $T=32$. Pauli measurement angle was $0.7$ radians relative to the quantization axis. This was done to ensure that the success of the protocol is not an artifact of making Pauli measurements along an axis where the time crystal ordering is expected.
\label{fig:supp_DTC}}
\end{figure*}

In this section, we provide some of the features of the interest function $f(U)$ and some observations on the behavior of the optimization iterations. To gain some insights into the behavior of $f$ as the number of qubits and number of time steps are varied, we estimate the function for unitaries picked from the DTC phase and trivial phase. Figure~\ref{fig:supp_DTC}(A) shows the dependence of $f$ on the number of time steps over which the states were picked. $f$ was averaged over 100 randomly chosen initial states and randomly chosen unitaries of the form in Eq.~\ref{eq:parametricUnitary_DTC} with $\langle J \rangle=\pi/4$, $\langle h \rangle=\pi/2$, $\hat{s}=\hat{z}$, $\hat{m}=\hat{x}$, and spatial disorder in $J,h$ picked uniformly randomly from $[-0.3,0.3]$. States for classification were collected from the time step $1$ to time step $T$. 
As a function of $T$ we empirically find the scaling $f\sim \sqrt{T}$ with statistical fluctuations that increase with $T$. Interestingly, $f$ is independent of number of sites $n$. 

For unitaries in the trivial phase the interest function weakly increases with $T$ but remains much smaller than in the DTC phase. Moreover the interest function appears to weakly decrease with $n$ in this case (Fig.~\ref{fig:supp_DTC}(B)). 

During the optimization, the system learns several features that favor the DTC phase. Figure~\ref{fig:supp_DTC}(C) shows the relative orientation of the direction of the Ising interaction and the field. The optimizer steers the two directions to be perpendicular to each other. 
Typically, this relative orientation of the Ising and the field direction is the feature that is learned first by the optimizer. 

In addition, the DTC oscillations are more robust if the direction of the Ising interaction aligns with the quantization axis of the computational basis states. The optimizer, consistent with this, steers $\hat{s}$ towards $\pm \hat{z}$ (while maintaining $\hat{m}$ perpendicular to $\hat{s}$). Other features like the value of $h_i$ and $J$ are learned later.

Figure~\ref{fig:supp_DTC}E shows examples of spatial profiles of the local fields $h_i$ that were discovered by the system at the end of the optimization. Majority of the sites have $h_i$ close to the optimal value with small disorder. In a small number of sites the field is close to $0$. We believe these configurations are local extrema of the interest function landscape. The belief is consistent with the observation in Fig.~\ref{fig:supp_DTC}(F,G) that the distribution of $h_i$ values show a small peak near $h_i=0$ whose height decreases with increase in the simplex size (learning rate) of the Nelder-Mead iterations. 

Lastly, we emphasize that the Pauli measurements for construction of the shadow representation can be performed along any three orthogonal directions. The optimal DTC phase has spin glass order along the  $z$ direction. It is useful to check that the success of the optimization protocol is not an accidental consequence of the Pauli-measurement scheme involving the $Z$-measurements - measurement scheme is agnostic to the DTC physics. To illustrate this, the Pauli measurements were made along a frame rotated relative to the qubit's quantization axis by $0.7$ radians in all examples.

\begin{figure*}
\includegraphics[width=1\textwidth]{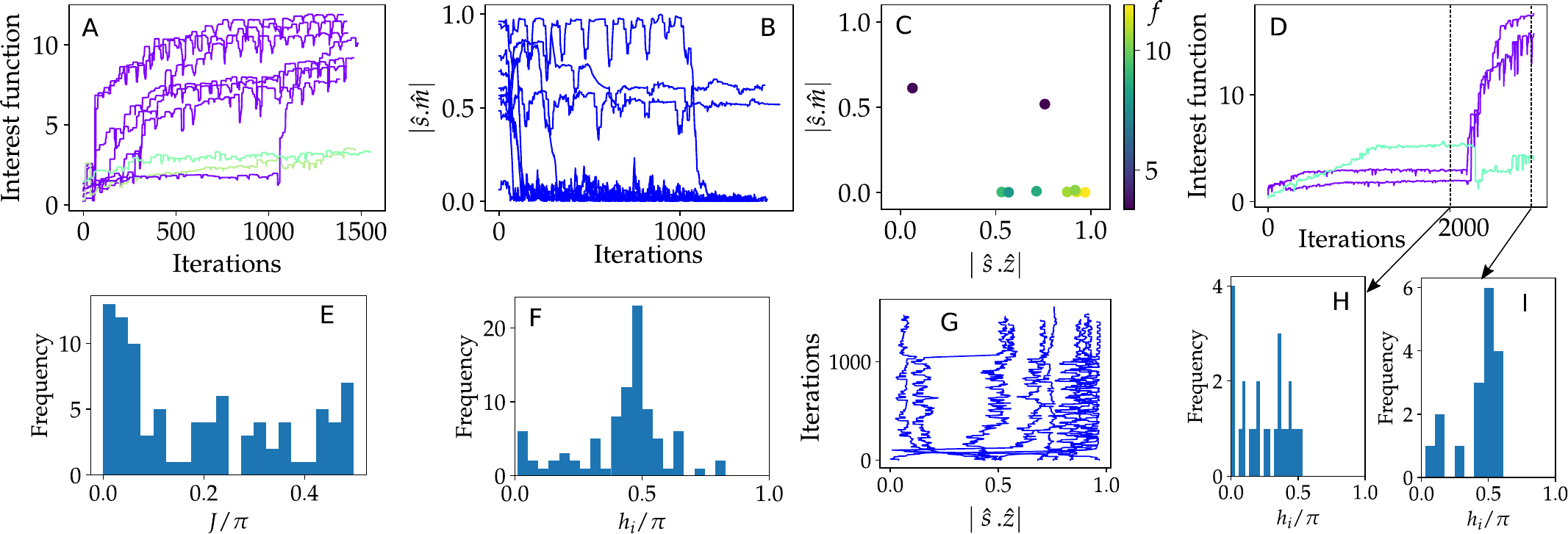}
\caption{(A) Evolution of interest function under a protocol where both $J_i$ and $h_i$ are site dependent. The lines corresponding to different optimization runs are colored based on the value of $|\hat{s}.\hat{m}|$ in the final configuration (B) Evolution of $|\hat{s}.\hat{m}|$, i.e. the inner product between directions of the Ising interactions and the fields, with iterations. (C) Final values of the $|\hat{m}.\hat{z}|$ and $|\hat{m}.\hat{s}|$ for different optimization trials. Color indicates the interest function of the final configuration. (G) Evolution of $|\hat{m}.\hat{z}|$ as a function of iterations. Panels (B,G) were aligned to match with the axes of Panel (C). (E) Distribution of $J_i$s in the final configuration. (F) Distribution of $h_i$ in the final optimized configuration. (D) Evolution of interest function with iteration in a protocol where the initial state $\psi$ is taken to be the maximally polarized state. For the first 2000 iterations the states used to estimate $f$ was taken from time steps $1$ to $32$. As this was unsuccessful, this was increased to $1$ to $100$. Panels (H,I) show the distribution of $h_i$ at the end of 2000 iterations and at the end of $\sim 2900$ iterations respectively. The results presented in panels A to C and E to G are generated with $n=10$ qubits and hyperparameters $N_s=500, N_{\rm init}=32, t_1=10, T=32$ and Pauli measurement angle $0.7$ radians.\label{fig:supp_DTC2}}
\end{figure*}

For a limited number of cases we ran the optimizations for a more general set of unitaries in which the couplings $J_i$ are position dependent:
\begin{align}
&U(\{J_i\}_{i=1,\dots n},\hat{s},\{h_i\}_{i=1,\dots n},\hat{m}) = e^{\imath H_J} e^{\imath H_h}\label{eq:parametricUnitary_DTC_sm}\\
&H_J=\sum_{i=1}^n J_i (\sigma_i.\hat{s})(\sigma_{i+1}.\hat{s}),\;\;H_h=\sum_{i=1}^n h_i \sigma_i.\hat{m}\nonumber
\end{align}
Other aspects of the optimization protocol remain the same. We have used periodic BCS in this calculation. The results summarized in Fig.~\ref{fig:supp_DTC2} A-C and E-G are qualitatively the same as in Fig.~\ref{fig:supp_DTC}; the optimization requires about twice as many iterations for convergence. As shown in panels B,G and C the features $\hat{s}.\hat{m}=0$ and $\hat{m}.\hat{z}\approx 1$ are learned very early in the iterations. Optimization of $h_i$ takes much longer. Within the hyperparameters used for the optimization $J_i$ had a broad distribution and was difficult to optimize. Final configurations had $J_i$ distribution with peaks near $J_i=0,\pi/2$ and around $\pi/4$.  

In all results presented so far in this section, the initial computational basis states were picked uniformly randomly. We also considered optimization of the unitaries while choosing the initial state to be the maximally polarized state. We found that the gradients in the interest function were small in this case unless the set of states $S$ used to compute $f$ were picked from very late-time steps. The observations from these tests are presented in Panels (D) and (H), (I). The first 2000 iterations of optimization were attempted with states taken from time steps $1$ to $32$. The optimizer was unable to improve the interest function even after 2000 iterations. The optimization procedure---more precisely the procedure for estimating $f$---was changed at this stage to use the first hundred time steps to collect the states for $f$ estimation. The optimizer worked better in this case, as can be seen in the distribution of $h_i$ extracted after 2000 and 2900 steps.

\section{Additional details for the interest landscape around dual-unitary circuits\label{App:DualU}}

\begin{figure*}
    \centering
    \includegraphics[width=\textwidth]{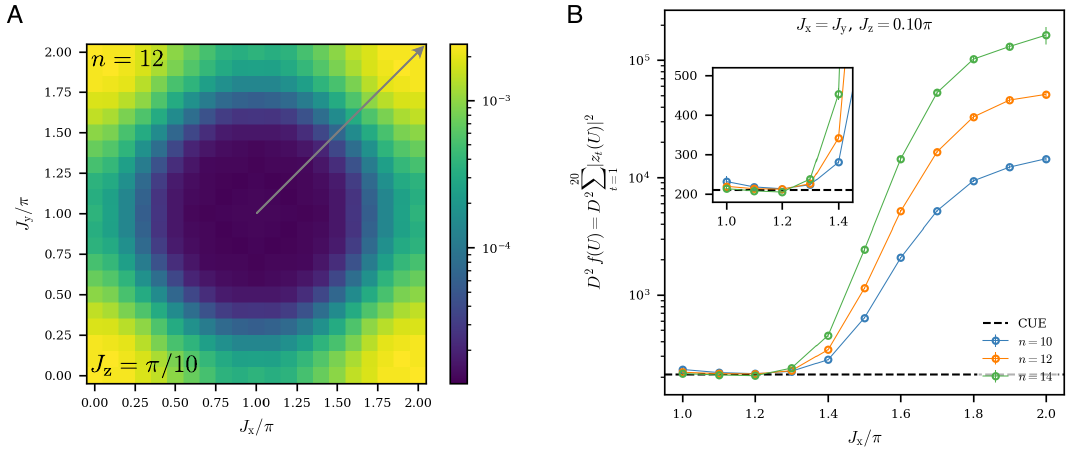}
    \caption{Extended data for the interest landscape close to dual-unitarity, for an interest function based on the spectral form factor as discussed in \autoref{sec:discover_dual_unitaries} of the main text.
    In particular we show in panel A (again) the interest landscape for the circuit discussed in the main text in the plane $J_{\rm z} = \pi/10$. And in panel B the (rescaled) interest function for a cut through parameter space, indicated by a gray arrow in panel A, for different system sizes $n$.}
    \label{suppl:fig:extended_sff}
\end{figure*}

In \autoref{suppl:fig:extended_sff} we show some additional data on the interest landscape around dual-unitary circuits, where $f(U) = - \sum_{t=1}^{t_{\rm max}} \overline{|z_t(U)|^2}$ as discussed in the main text and we set $t_{\rm max} = 20$. The data shown in panel A is identical to that shown in panel E of Fig. 4 in the main text. In panel B, we show in addition the interest function along a diagonal cut ($J_{\rm x} = J_{\rm y}$) for different system sizes. 

Even for the modest system sizes considered here, the gradient only becomes unresolvable quite close to the dual-unitary point, at $|J_{\rm x}-\pi| \approx 0.2\pi$. Interestingly however, for $|J_{\rm x} - \pi | \lesssim  0.2\pi$ the gradient has no visible system size dependence and seems even to decrease with system size. This behavior is somewhat unexpected, as it indicates a surprising stability of dual-unitarity to perturbations of the circuit.
For the purpose of the program outlined in this paper, this has two possible implications. Either these vanishing gradients are simply a finite size effect, in which case on a scalable quantum computer one would be able to optimize much closer to the dual-unitary point. 
The even more exciting possibility would be that the extremal spectral repulsion exhibited by dual-unitary circuits is more stable than previously imagined.
Either way, we believe this example shows that the study of interest landscapes and their maxima is worthwhile and leads to interesting questions or discoveries.

\section{Measuring the Spectral Form Factor\label{app:SFF}}

In this Appendix, we discuss two possible routes towards experimental realization of our proposed protocol to discover maximally chaotic finite-depth Floquet circuits.

\subsection{Hadamard test}

\begin{figure}[h]
\centering

\begin{quantikz}[column sep=0.45cm]
\lstick{$\ket{0}$}         & \gate{H} & \ctrl{1}      & \meter{\!\langle X\rangle, \!\langle Y\rangle} \\
\lstick{$\rho=\mathbb{1}/2^n$} & \qw      & \gate{U^{t}}  & \qw \\
\end{quantikz}

\caption{Controlled-unitary based circuit estimating $z_t(U)$. The measured expectation values on the ancilla are $\expval{X} = {\rm Re}\left(z_t(U)\right)$ and $\expval{Y} =  {\rm Im}\left(z_t(U)\right)$.}
\label{fig:hadamart_test_sff}

\end{figure}

An elementary method to obtain the spectral form factor is by performing a Hadamard test, that is estimating the quantity $z_t(U)$ with a simple circuit applying a controlled $U^t$ operation (see, e.g., Ref. \onlinecite{knill1998DQC1}). We sketch the relevant circuit in \autoref{fig:hadamart_test_sff}. 

However,  this naive method suffers from an exponentially small signal-to-noise ratio in general.
To see this, note that after the circuit in \autoref{fig:hadamart_test_sff}, the expectation values on the ancilla are given by
\begin{align}
    \expval{X} &=  \Re\left(z_t(U)\right), \\
    \expval{Y} &=  \Im\left(z_t(U)\right),\\
    \expval{Z} &= 0,
\end{align}
i.e., $z_t(U)=\langle X\rangle+i\langle Y\rangle$. However, in order to distinguish the random-matrix-theory prediction $|z_t(U)|^2=t 2^{-2n}$, realized by dual-unitary circuits, from the behavior $|z_t(U)|^2 \approx 2^{-n}$ of generic chaotic circuits, the sample complexity increases exponentially with the number of qubits $n$.
A slight adaptation of the Hadamard test protocol to multiple ancillas, as well as a detailed discussion of its experimental realization and resource estimates can be found in Ref. \onlinecite{vasilyev2020sff}.

\subsection{Randomized measurements and partial spectral form factor}

\begin{figure}
    \centering
    \includegraphics{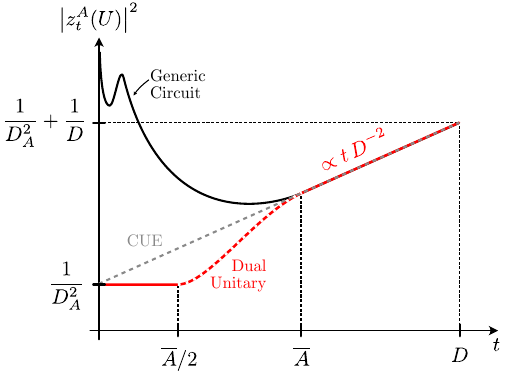}
    \caption{Sketch of the behavior of the partial spectral form factor [\autoref{eq:psff}] as function of time in generic circuits, the circular unitary ensemble, and dual-unitary circuits. For $\abs{A} = n$ (i.e. $\abs{\overline A} = 0$) the behavior of the full spectral form factor in \autoref{fig:discover_dual_unitaries} C is recovered. Note that for all $\abs{A} < n$, minimization of the sum of $\abs{z_t^A(U)}^2$ over time steers the system towards dual-unitary circuits.}
    \label{fig:pSFF_sketch}
\end{figure}

A more practical approach to quantify spectral signatures of quantum many-body chaos is the measurement of the so-called partial spectral form factor (pSFF) \cite{gong2020pSFF,joshi2022pSFF}. Measurement of this quantity has already been realized on current hardware \cite{Collins2023sff_experiment, dong2025sff_experiment}, and, most importantly, it is efficiently scalable to larger systems. For a given subsystem $A$, the pSFF is defined as\footnote{Note that here $z_{t}^A(U)=\frac{\tr_{A}(U^t)}{\tr_{A}\mathbb{1}_A}$ is an operator acting on $\bar A$, and---with a little abuse of notation---we denoted by $|O|=\sqrt{\tr (O^\dagger O)/\tr \mathbb{1}}$ the Hilbert-Schmidt norm.} 
\begin{equation}\label{eq:psff}
    |z_{t}^A(U)|^2 = \frac{1}{D_{\bar A}} \frac{1}{D^2_{ A}}\tr_{\bar{A}}\left[ \tr_{A}(U^t)^\dagger\tr_{A}(U^t)\right]
\end{equation}
where $\bar A$ is the complementary subsystem to $A$, $\tr_{A}$ ($\tr_{\bar A}$) denotes the partial trace over the qubits in $A$ (respectively, in $\bar A$), and $D_A=2^{\abs{A}}$ ($D_{\bar A}=2^{\abs{\bar A}}$) is the Hilbert-space dimension of the subsystem $A$ ($\bar A$), with $|A|+|\bar A|=n$; in the following, we will assume $1\ll |A| \ll n$. The full SFF is obtained in the limit where $A$ grows to the entire system. For finite subsystem size, the pSFF differs from the SFF in general, but is still expected to be a witness of quantum many-body chaos. For example, it was shown in Ref. \onlinecite{joshi2022pSFF} that for fully random unitaries in $U(2^n)$, the partial spectral form factor follows a `shifted ramp plateau' behavior, 
\begin{eqnarray}
    \overline{|z_{t}^A(U)|^2} \sim 
    \frac 1 {D_{A}^{2}} + \frac{t}{D^2} \, .
\end{eqnarray}
The pSFF of generic finite-depth unitary circuits, instead, exhibits a more structured behavior at initial times than the SFF, consisting of a rapid drop, followed by a large asymmetric peak, eventually  followed by a (shifted) universal ramp after a time scale of order $|\bar A|$ \cite{yoshimura2025dynamics}.

While dual-unitary circuits realize the exact random-matrix-theory SFF $\overline{|z_t(U)|^2}=t/D^2$ for all $t$, they do not do the same for the pSFF. In particular, it is straightforward to see that dual-unitary circuits   have a flat pSFF at initial times, $|z^A_t(U)|^2=1/D_A^2$ for $t<|\bar A|/2$~\cite{fritzsch2024pSFF}. This behavior is followed by a rise and eventually matches the shifted ramp plateau behavior~\cite{fritzsch2024pSFF}:
\begin{align}
    |z^A_t(U)|^2 = \begin{cases}
        1/D_A^{2} & \text{for}~ t < |\bar A| / 2 \\
        1/D_A^{2}+t/D^2 & \text{for}~ t \gtrsim |\bar A|
    \end{cases}
\end{align}

This result, illustrated in \autoref{fig:pSFF_sketch}, shows that minimization of the pSFF is expected to be an excellent  proxy for identifying special maximally chaotic finite-depth unitary circuits. 

\begin{figure}
    \centering
\begin{quantikz}[row sep={0.6cm,between origins}, column sep=0.5cm]
\lstick{$\ket{0}$}  & \gate{u_1} & \qw & \gate[4, nwires=3]{U^t} & \qw & \gate{u_1^{\dagger}} & \meter{$Z$} \\[0.5em]
\lstick{$\ket{0}$}  & \gate{u_2} & \qw &                         & \qw & \gate{u_2^{\dagger}} & \meter{} \\
\lstick{\vdots} \setwiretype{n} & \vdots & &                     &     & \vdots               &  \\[0.5em]
\lstick{$\ket{0}$}  & \gate{u_n} & \qw &                         & \qw & \gate{u_N^{\dagger}} & \meter{}
\end{quantikz}

    \caption{Randomized measurement protocol for estimating the (partial) spectral form factor. The single-qubit unitaries $u_1, \dots u_n$ are chosen from some unitary 2-design of the Haar ensemble. The (partial) spectral form factor can then be estimated using  \autoref{eq:psff_randomized}.}
    \label{fig:psff_randomized}
\end{figure}

Crucially, randomized measurements \cite{Aaronson2020, Huang2020Predicting,Elben2018Renyi,elben2023review, joshi2022pSFF} allow estimation of the pSFF with a sample complexity which is only exponential in \emph{subsystem} size $\abs{A}$. In particular, as proposed in Ref. \onlinecite{joshi2022pSFF}, we can create a random product state, apply the unitary $U^t$ and then measure in the conjugate random product basis, as sketched in \autoref{fig:psff_randomized}. Each measurement yields a bit string outcome $\vec s \in \{0, 1\}^n$, and after $M$ measurements we have a collection of outcomes $\{\vec s^{(r)}\}_{r = 1}^M$, from which the pSFF can be estimated as
\begin{equation}\label{eq:psff_randomized}
    |z^A_t(U)|^2 =  \frac{2^{n + \abs{A}}}{M} \sum_r^{M} (-2)^{-\abs{\vec s_A^{(r)}}}
\end{equation}
where $\vec s_{A} = (s_i)_{i\in A}$ is the restriction of the measurement outcomes to the subsystem $A$, and $\abs{\vec x} = \sum_i x_i$ denotes the Hamming weight.
The typical sample complexity of the above procedure to access the pSFF of a given subsystem $A$ up to fixed relative error is expected to be $M \sim 10^{\abs{A}} \approx 2^{3.32 \abs{A}}$.
We note that in practice, one can combine the averaging over randomized measurements in \autoref{eq:psff_randomized} with the average over a given ensemble of circuits (i.e., it suffices to perform a 'single-shot' averaging over realizations).

In summary, although in large systems one is in practice restricted to compute the pSFF for small subsystems $A$, given the discussion above this suffices to distinguish generic chaotic Floquet circuits from special circuit instances with extremal spectral correlations, such as dual-unitary circuits. We thus expect our discovery protocol to be experimentally realistic with currently available resources.

\end{document}